\documentclass[11pt,a4paper]{article}
\pdfoutput=1
\usepackage{jheppub} 
\usepackage{tikz}
\usetikzlibrary{decorations.pathmorphing}

\usepackage{amsmath,amssymb,amsbsy,amsfonts,latexsym,graphicx}
\usepackage{hyperref}
\usepackage{color,array,subfigure}

\def\D  {\Delta}       \def\e  {\epsilon}

\renewcommand{\a}{\alpha}      \renewcommand{\b}{\beta}
      
     \renewcommand{\L}{\Lambda}
      \renewcommand{\t}{\tau}
\renewcommand{\th}{\theta}

 \newcommand{\w}{\omega}

\newcommand{\be}{\begin{equation}}
	\newcommand{\ee}{\end{equation}}
\newcommand{\bea}{\begin{eqnarray}}
	\newcommand{\eea}{\end{eqnarray}}
\newcommand{\nn}{\nonumber}

\newcommand{\ra}{\rightarrow}

\allowdisplaybreaks

\def\w{{\omega}}

\def\t{{\tau}}
\def\a{{\alpha}}

\def\D{{\Delta}}

\def\b{{\beta}}

\def\L{{\Lambda}}

\def\p{\partial}
\def\f {\frac}

\def\nn{\nonumber}

\title{Islands in Proliferating de Sitter Spaces}
\author[a]{Jong-Hyun Baek}
\author[a,b]{and Kang-Sin Choi}

\affiliation[a]{Institute of Mathematical Sciences, Ewha Womans University, Seoul 03760, Korea}
\affiliation[b]{Scranton Honors Program, Ewha Womans University, Seoul 03760, Korea}
\emailAdd{jonghbaek@gmail.com}
\emailAdd{kangsin@ewha.ac.kr}

\abstract{We study two-dimensional de Sitter universe which evolves and proliferates according to the Ginsparg--Perry--Bousso--Hawking mechanism, using Jackiw--Teitelboim gravity coupled to conformal matter. Black holes are generated by quantum gravity effects from pure de Sitter space and then evaporate to yield multiple disjoint de Sitter spaces. The back-reaction from the matter CFT is taken into account for the dilaton as a function of the temperature of the CFT. We discuss the evaporation of black holes and calculate the finite temperature entropy of an inflating region using the island formula. We find that the island moves towards the apparent horizon of the black hole as the temperature increases. The results are applied to the case of  multiple evaporating black holes, for which we suggest multiple islands.}


\begin{document}
	\maketitle
	\flushbottom

\section{Introduction}

De Sitter (dS) space is a key ingredient to understand the history of the universe. It can be a simple model of the inflationary phase in the very early universe. 
Similar to a black hole, de Sitter space possesses a cosmological horizon and has a characteristic temperature \cite{Gibbons:1977mu}. An observer in the static patch sees a thermal equilibrium at the dS temperature. 

Another aspect of interest is that de Sitter space has a semiclassical instability, leading to the nucleation of black holes \cite{Ginsparg:1982rs}. A gravitational instanton effect can generate a Schwarzschild black hole, which is extremal in size and can evaporate subsequently.   
This black hole generating process can be regarded as a spontaneous topology change, where a spatial section of de Sitter space $S^3$ undergoes a transition to $S^1 \times S^2$ according to the Ginsparg--Perry--Bousso--Hawking (GPBH) mechanism \cite{Ginsparg:1982rs,Bousso:1997wi}. The latter geometry is described by the Schwarzschild-de Sitter (SdS) solution \cite{Kottler}  in the Nariai limit \cite{Nariai}. For evaporation of black holes, we may see this geometry as an $S^2$ fibration over $S^1$, that is, the radius of $S^2$ is not uniform along the $S^1$.
As the black holes evaporate, the $S^2$ factor shrinks to points at the black hole locations, pinching off the total space to a number of $S^3$'s corresponding to pure de Sitter spaces \cite{Bousso:1998bn,Bousso:2002fq}.  

In this paper we study this process in two-dimensions by utilizing Jackiw--Teitelboim (JT) gravity \cite{Teitelboim:1983ux, Jackiw:1984je}. In JT gravity, the essential features of higher-dimensional theory is captured by the dilaton, which can be roughly identified with the radius of $S^2$ in this case. Thus fluctuations of the dilaton describe deviation from the Nariai limit, where the black hole horizon has the same size as the cosmological one \cite{Maldacena:2019cbz, Cotler:2019nbi}. 

We couple a conformal matter to gravity and consider an equilibrium state with the black hole at finite temperature. The backreaction from the semiclassical energy momentum tensor of the matter CFT changes the causal structure of the geometry and correspondingly the Penrose diagram.  We use the backreacted dilaton solution and impose the boundary condition such that the dilaton approaches the one for pure de Sitter at future boundary and calculate the entanglement entropy of a non-gravitating region in the context of evaporating black holes. 

The island formula \cite{Penington:2019npb,Almheiri:2019psf, Almheiri:2019hni} computes the von Neumann entropy of system $R$ through
\begin{equation} \label{vNentropy}
	S(\rho_R) = \text{min ext}_I\, S_{\text{gen}}(I \cup R),
\end{equation}
where the generalized entropy is given by
\begin{equation}
	S_{\text{gen}}(I\cup R) = \f{A(\p I)}{4G_N}+S_{\text{mat}}(I \cup R).
\end{equation}
Here the island $I$ is a gravitating region and  $A(\p I)$ is the area of the boundary of $I$ and $S_{\text{mat}}(I\cup R)$ is the von Neunmann entropy of the system $I\cup R$. The generalized entropy is extremized over the configurations of $I$ and then minimized over the choices. The island rule was applied in the eternal black hole in \cite{Almheiri:2019yqk} and was derived using the semiclassical gravitational path integral with the replica method \cite{Penington:2019kki, Almheiri:2019qdq, Almheiri:2020cfm}.   

We study the entanglement entropy of the newly generated de Sitter space, which can be seen as follows.
An observer in the ``Milne'' patch (an inflating region near the spacelike future infinity $\cal{I}^+$, see Fig. \ref{fig:eternalds2}) can collect the radiation from the evaporating black hole. The matter entropy increases as the CFT temperature increases until the island appears purifying the entangled pairs of radiation. 
Above the Page temperature, the generalized entropy becomes relevant and it decreases over temperature. 
This shows that the final dS space after evaporation is a pure state. 
(It is not necessary to consider dS space as the final state. Minkowski space(s) are considered in Refs. \cite{Chen:2020tes,Hartman:2020khs} in different contexts.)
	
In Section 2, we review the JT de Sitter gravity and its origin from the four-dimensional Einstein gravity. Higher dimensional Nariai geometry is reduced to two-dimensions to obtain the global de Sitter space. In Section 3, we explain the backreacted dilaton solution and the boundary condition at future infinity. The changes in causal structure and the Penrose diagrams due to backreaction are described. We also discuss how evaporation of black hole in Bousso--Hawking mechanism can be modeled with the solution. In Section 4, the entropy of an inflating region is calculated using the island formula. High, low and close-to-the-critical temperature limits are considered and numerical method is used in the intermediate temperature regime. In Section 5, the nucleation of multiple black holes are considered and attaching the new de Sitter spaces at future boundary is discussed. The entropy formula is generalized to incorporate the multiple cover of dS and the change in the Casimir energy. The computation of the entanglement entropy of multiple regions shows that each new de Sitter space is pure. In Section 6, we conclude with a summary and future directions.

\section{JT gravity in de Sitter space}

We briefly review the de Sitter JT gravity.
The four-dimensional Schwarzschild-de Sitter black hole is described by the metric,
\begin{equation}
 ds^2 = - f(\tilde{r}) d\tilde{t}^2 + f(\tilde{r})^{-1} d\tilde{r}^2 + \tilde{r}^2 d \Omega^2, \qquad f(\tilde{r})=1-\frac{2\mu}{\tilde{r}}-\frac{\L}{3}\tilde{r}^2,
\end{equation}
where $\mu$ is a mass parameter and $\L$ is a cosmological constant. If $\mu=0$, the metric reduces to pure de Sitter space in the static patch. The function $f(\tilde{r})$ has two positive roots for $0<\mu<\f{1}{3\sqrt{\L}}$, one of them corresponding to the black hole horizon  and the other to the cosmological horizon. When the two roots coincide, the geometry is described by the Nariai solution \cite{Nariai}. 

In the near-Nariai limit of the SdS black hole, the metric takes the factorized form of dS$_2 \times S^2$,
\begin{equation}
 ds^2 = -\left(1-\f{r^2}{L^2}\right)dt^2 + \left(1-\f{r^2}{L^2}\right)^{-1}dr^2 + L^2\phi\, d\Omega_2^2,
\end{equation}
where the $L$ is the radius of $dS_2$ and $\phi$ is the dilaton, which can be written as 
\begin{equation} \label{dilaton}
 \phi = \phi_0 + \Phi.
\end{equation}
The constant $\phi_0$ corresponds to the radius of the $S^2$ in the Nariai limit, and $\Phi$ to fluctuation away from it. 
 
Dimensional reduction of Einstein gravity with a positive cosmological constant along the $S^2$ gives the action of de Sitter JT gravity \cite{Maldacena:2019cbz, Cotler:2019nbi, Svesko:2022txo}.
\begin{equation}
	I =\f{\phi_0}{16\pi G_N}\int d^2x\sqrt{-g}R + \f{1}{16\pi G_N}\int d^2x\sqrt{-g}\Phi\left(R-\f{2}{L^2}\right) + I_{\text{CFT}}[g], 
\end{equation}
where the boundary terms are omitted, and $I_{\text{CFT}}$ is the action of the matter CFT, which couples to the metric. Here $G_N$ is the two-dimensional Newton's constant related to the four-dimensional one, $G_N =  G_N^{(4)}/(4\pi L^2)$.

The dilaton constrains the metric to dS$_2$. In conformal gauge and lightcone coordinates, $x^{\pm}=L(\t\pm\theta)$, the metric is written 
\begin{equation}\label{dsmet}
	ds^2=\f{L^2}{\cos\t^2}(-d\t^2+d\th^2)=-e^{2\w}dx^+dx^-, \qquad e^{-2\w}=\cos^2\t,
\end{equation}
where $\t\in(-\pi/2,\pi/2)$ and $\th\in(-\pi/2,3\pi/2)$. The metric equations of motion are given by
\begin{equation}\label{eom1}
	e^{2\w}\p_{\pm}\Big[e^{-2\w}\p_{\pm}\Phi\Big]=-8\pi G_N\langle T_{\pm\pm} \rangle,
\end{equation}
and 
\begin{equation}\label{eom2}
	2\p_+\p_-\Phi -\f{1}{L^2}e^{2\w}\Phi   =16\pi G_N\langle T_{+-}\rangle,
\end{equation}
where $\langle T_{ab}\rangle$ is the expectation value of the stress tensor of the matter CFT.

\begin{figure} \begin{center}
		\begin{tikzpicture}
			\draw (0,0) -- (0,4);
			\draw [decorate,decoration={zigzag,segment length=2.2mm, amplitude=0.6mm}] (0,4) -- (4,4)
			node[midway, below=0.5cm] {BH};
			\draw (0,4) node[left, inner sep=2mm] {$\tau=\frac{\pi}{2}$};
			\draw [decorate,decoration={zigzag,segment length=2.2mm, amplitude=0.6mm}] (0,0) -- (4,0);
			\draw (0,0) node[ left, inner sep=2mm] {$\tau=-\frac{\pi}{2}$};
			\draw (4,0) -- (8,0) -- (8,4);
			\draw (4,4) -- (8,4) node[midway, below=0.5cm] {Milne};
			\draw (4,4) -- (4,0);
			\draw (0,0) -- (4,4);
			\draw (4,0) -- (0,4);
			\draw (8,0) -- (4,4);
			\draw (8,4) -- (4,0);
			\draw (6,0.1) -- (6,-0.1) node [below] {0};
			\draw (4,0.1) -- (4,-0.1) node[below] {$\f{\pi}{2}$};
			\draw (8,0.1)--(8,-0.1) node[below] {-$\f{\pi}{2}$};
			\draw (2,0.1)--(2,-0.1) node [below] {$\pi$};
			\filldraw[black] (2,2) circle (2pt);
			\filldraw[blue] (6,2) circle (2pt);
			\draw (0,2) node[sloped, rotate=-50] {$\parallel$};
			\draw (8,2) node[sloped, rotate=-50] {$\parallel$};
			\draw (6,4) node[above] {$\cal{I}^+$};
		\end{tikzpicture}
	\end{center} 
	\caption{Penrose diagram for the two-dimensional de Sitter space. Singularities are at $\tau=\pm \pi/2$ and $\cos \theta <0.$ The black and blue dots are the black hole horizon and cosmological horizon, respectively. Two sides are indentified.}
	\label{fig:eternalds2}
\end{figure}
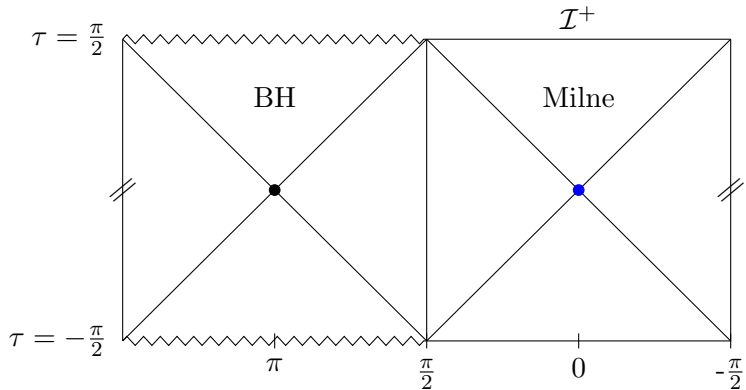

In the absence of the energy--momentum tensor, the sourceless dilaton $\Phi$ is given by 
\begin{equation}\label{dilaton1}
 \Phi_{\text{sl}}(\t,\th) =\bar{\phi}L \frac{\cos \theta}{\cos \tau} = \bar{\phi}r,
\end{equation}
for a constant $\bar{\phi}$ and it has negative values for $\pi/2<\th<3\pi/2$. Near future infinity $\t=\pi/2$, the dilaton goes to $-\infty$ in the black hole or the crunching region and to $+\infty$ in the inflating region with $|\th|<\pi/2$. This solution describes two-dimensional de Sitter space obtained from SdS$_4$ and drawn in  \autoref{fig:eternalds2}. Although this two-dimensional dS space has no singularity, the dilaton $\Phi$ at $r=-\infty$ indicates the existence of singularity in the four dimensional black hole at the corresponding region.

There is a different way to obtain dS$_2$, that is a circle reduction from dS$_3$. In this case, the range of the $\th$ coordinate is a segment, half in size with no identification of both ends. This dS$_2$ contains no singularity and the dilaton is always positive since $-\pi/2 \leq \th \leq \pi/2$. It's called the half reduction, while the above dS$_2$ from SdS$_4$ is called the full reduction in \cite{Svesko:2022txo}.   

From the two-dimensional point of view, the nucleation of black holes from pure dS$_2$ and subsequent evaporation back into a number of pure de Sitter spaces with no black hole can be understood as a transition in the $\th$ coordinate from a segment to a circle and then into a number of segments. As black holes evaporate the dilaton $\Phi$ goes to minus infinity at the black hole horizon, pinching off the circle into segments. 

 In \cite{Bousso:1998bn}, this process was described using the perturbation of the radius in the transverse $S^2$. The perturbation changes along the $\th$ direction of dS$_2$ and can be decomposed into Fourier modes with mode number $n$. More than one black holes can nucleate if a higher fluctuation $n>1$ mode of the perturbation dominates, with probability $e^{-n^2}$. It may be that de Sitter spaces have proliferated by the mechanism before and during inflationary phases, giving rise to eternal inflation.

\section{Solutions with backreaction from CFT} \label{sec:backreactedsol}

In this section, we review the solution of Balasubramanian, Kar and Ugajin (BKU) \cite{Balasubramanian:2020xqf} with a slight generalization of different left, right temperatures. It describes  de Sitter space with a black hole in thermal equilibrium with the matter CFT at fixed temperature $T = 1/(\beta L)$. The backreaction from the matter deforms the causal structure of the spacetime and the Penrose diagram. We explain how the black hole evaporation can be understood as we increase the temperature. We mainly focus on the hyperbolic patches of de Sitter space but for studies in the static patch, see \cite{Sybesma:2020fxg, Aalsma:2021bit, Kames-King:2021etp, Aalsma:2022swk}.

\subsection{Backreacted dilaton} \label{sec:dilaton}

We consider a thermal state on a flat space such that the stress tensor is of the form
\begin{equation}\label{flatt}
	\t_{\pm\pm} = \f{c}{48\pi L^2}\left(\f{2\pi}{\b_{\pm}}\right)^2 - \f{c}{48\pi L^2} ,
\end{equation}
where $c$ is the central charge of the CFT and $\b_{\pm}$ are the left/right inverse temperature and the second term is the Casimir energy due to the periodicity of $\theta \simeq \theta + 2\pi$.  They are to be further discussed in Sections \ref{sec:fintempentropy} and \ref{sec:casimir}.

The expectation value of the stress tensor on a curved space can be obtained by using the trace anomaly and the conservation equation of stress tensor, 
\begin{equation} \begin{split}
	\langle T_{\pm\pm} \rangle  &= \f{c}{12\pi}\Big(\p_{\pm}^2\w-(\p_{\pm}\w)^2\Big) +\t_{\pm\pm}, \\
	\langle T_{+-} \rangle& = -\f{c}{12\pi}\p_+\p_-\w,
\end{split}
\end{equation}
where $\w$ is the exponent of the conformal factor in \eqref{dsmet}.
For the state \eqref{flatt}, we have
\begin{equation}
	\langle T_{\pm\pm} \rangle  = \f{c}{48\pi L^2}\left(\f{2\pi}{\b_{\pm}}\right)^2, \qquad \langle T_{+-} \rangle = -\f{c}{48\pi L^2\cos^2\t}.
\end{equation}
The Casimir energy is canceled by the contribution from the warp factor $\omega$. These are going to describe the thermal radiation from the black hole. Note that the temperature $1/(\beta_{\pm}L)$ is different from the Gibbons--Hawking temperture of de Sitter space, $T_{\rm GH}=1/(2\pi L)$.

With the stress tensor expectation value, a solution to \eqref{eom1} and \eqref{eom2} can be found
\begin{equation}\label{dilaton}
	\Phi(\t,\th) = \a\f{\cos\th}{\cos\t} - \f{1}{2}\left[\Big((K_++K_-)\t+ (K_+-K_-)\th\Big)\tan\t +K_++K_-\right] +\f{cG_N}{3},
\end{equation}
where 
\begin{equation}\label{Kpm}
	K_{\pm}=\f{cG_N}{3}\left(\f{2\pi}{\b_{\pm}}\right)^2.
\end{equation}

Since we are interested in the transition to pure dS$_2$ after the black hole has evaporated, we may impose the boundary condition at late time, $\t\ra\pi/2$, that the dilaton approaches the sourceless solution \eqref{dilaton1}, presumably in a new coordinate.
The effects of the backreaction can be seen as the $SO(1,1)$ boost on the coordinates $r$ and $t$ \cite{Balasubramanian:2020xqf},
\begin{equation}\label{boost}
	\f{r'}{L}=b_+\f{\cos\th}{\cos\t}  - b_-\tan\t, \qquad \tanh \f{t'}{L}=b_+\f{\sin\t}{\sin \th}-b_-\f{1}{\tan\th}
\end{equation} 
where the rotation parameter $b_{\pm}=(b\pm b^{-1})/2$ with $b \geq 1$.   This transformation is a subgroup of the isometry $SO(1,2)$ of dS$_2$.
Thus we expect the dilaton of the form  \cite{Balasubramanian:2020xqf},
\begin{eqnarray}\label{dilatonsl}
	\Phi_0(\t,\th) &=&\bar\phi r' = \bar{\phi}L\left[ b_+\f{\cos\th}{\cos\t} - b_-\tan\t \right].
\end{eqnarray}  
in the limit $b=1$, this solution is simplified to \eqref{dilaton1}.

For $\tau \to \pi/2$, at leading order in $\tau$, the dilaton solution \eqref{dilaton} is expanded as
\begin{equation}
	\Phi(\t,\th) = \f{\f{\pi}{4}(K_++K_-)-\f{1}{2}(K_+-K_-)\th -\a\cos\th}{\t-\f{\pi}{2}}+ \cdots
\end{equation}
In order to match the leading order coefficients of $\Phi_0$ and $\Phi$, we need to take $K_+=K_-=K$, with $K$ defined as in \eqref{Kpm} but with $\b$. Then the parameters are related 
\begin{equation}\label{paramap}
	 K = \f{cG_N}{3}\left(\f{2\pi}{\b}\right)^2=\f{\bar{\phi}L}{\pi}\left(b-\f{1}{b}\right) ,\qquad \a = \f{\bar{\phi}L}{2}\left(b+\f{1}{b}\right).
\end{equation}
Thus the limit $b=1$ corresponds to the zero temperature limit $\beta \to \infty$.
The dilaton satisfying the boundary condition is given by
\begin{equation}\label{dilatonf}
\begin{split}
 	\phi&=\phi_0+\Phi(\t,\th)\\
	 &\equiv \phi_0+\f{\bar{\phi}L}{2}\left[\left(b+\f{1}{b}\right)\f{\cos\th}{\cos\t} - \f{2}{\pi}\left(b-\f{1}{b}\right)(\t\tan\t+1) \right] +\f{cG_N}{3}.
\end{split}
\end{equation}
Note that the dilaton has dependence on the temperature through the parameter $b$ by the relation \eqref{paramap}. As $b$ gets larger, the effect of backreaction becomes stronger. We assume $G_N \ll 1, c \gg 1$ with $c G_N$ fixed to have the semiclassical limit and $  c G_N \ll \bar \phi L \ll \phi_0$ to make the CFT contribution dominant \cite{Svesko:2022txo}. 

\subsection{Backreacted horizons}

The dilaton \eqref{dilatonf} defines the black hole and the cosmological apparent horizons by $\p_{\t}\phi=\p_{\th}\phi=0$. They are located at $(\t,\th)=(0,\pi)$ for black hole and $(\t_0,0)$ for cosmological apparent horizon at large enough $b$, satisfying
\begin{equation}
	\sin\t_0 = \f{b^2-1}{b^2+1}.
\end{equation}

\begin{figure} \begin{center}
\begin{tikzpicture}
\draw [gray, dashed] (0,0) -- (0,4);
      \draw[gray, dashed] (0,4) -- (4,4)
     ;
\draw[gray, dashed] (0,0) -- (4,0)
    ;
\draw [gray, dashed] (4,0) -- (8,0) -- (8,4) -- (4,4) -- (4,0);
\draw (4.5,0) -- (4.5,4);
\draw (-0.5,0) -- (-0.5,4);
\draw (7.5,0) -- (7.5,4);
\draw [decorate,decoration={zigzag,segment length=2.2mm, amplitude=0.6mm}] (-0.5,0) .. controls (2,1) .. (4.5,0);
\draw [decorate,decoration={zigzag,segment length=2.2mm, amplitude=0.6mm}] (-0.5,4) .. controls (2,3) .. (4.5,4);
\draw  (4.5,0) .. controls (6,0.5) .. (7.5,0);
\draw  (4.5,4) .. controls (6,3.5) .. (7.5,4);
\draw [very thick, blue] (4.8,3.8) .. controls (6,3.4) .. (7.2,3.8)
node[midway, below] {$R$};
\draw [very thick, magenta] (0.2,3.6) .. controls (2,2.7) .. (3.8,3.6)
node[midway, below] {$I$};
\draw (-0.5,0) -- (1.5,2) -- (-0.5,4);
\draw (4.5,0) -- (2.5,2) -- (4.5,4);
\draw (4.5,4) -- (6,2.5) -- (7.5,4);
\draw (4.5,0) -- (6,1.5) -- (7.5,0);
\filldraw[black] (2,2) circle (2pt);
\filldraw[red] (6,2.5) circle (2pt);
\filldraw[red] (6,1.5) circle (2pt);
\draw (2,0.1)--(2,-0.1) node [below] {$\pi$};
\draw (6,0.1)--(6,-0.1) node [below] {$0$};
\end{tikzpicture}
\end{center} 
\caption{Backreacted de Sitter at finite temperature \cite{Balasubramanian:2020xqf}. The dotted lines are Penrose diagram at zero temperature.  The black dot is the apparent horizon of the black hole and red dots are the cosmological horizons.}
\label{backreactedds}
\end{figure}
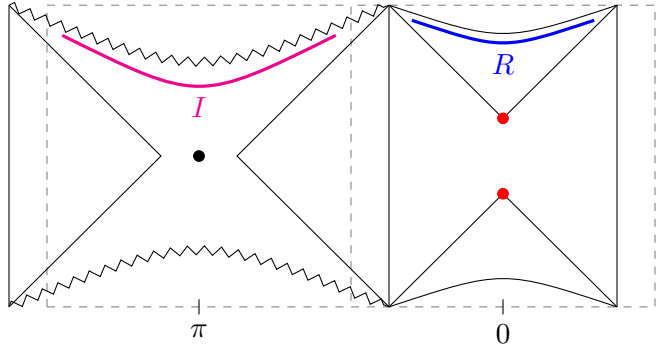

It was shown in \cite{Balasubramanian:2020xqf} that the dilaton at the black hole apparent horizon, $\Phi(0,\pi)$ in \eqref{dilatonf}, decreases as the temperature of the CFT increases. This was interpreted as the evaporation of the black hole.
The dilaton at the cosmological event horizon, which coincides with the cosmological apparent horizon at high temperature, reaches at the upper limit in the high temperature limit: $\phi(\t_0,0)|_{b=\infty}=\phi_0+\bar{\phi}L$. The backreaction also changes the Penrose diagram as in Fig.\,\ref{backreactedds}, where the future singularity is elongated and curves downward. This effect is due to the change in the range of $\th$ for which the dilaton $\Phi$ becomes negative as $\t \rightarrow \pi/2$. 

We can also see changes in the horizons by coordinate transformations. Before turning on the CFT temperature $\b$, the de Sitter metric and dilaton are written in the static patch 
\begin{equation}\label{Milne}
 ds^2 = \left(1-\frac{r^2}{L^2}\right) dt^2 +\left(1-\frac{r^2}{L^2}\right)^{-1} dr^2, \qquad \Phi = \bar{\phi}r.
\end{equation}
The cosmic horizon is at $r=L$ and the black hole horizon at $r=-L$. With the same metric, 
the Milne patch corresponds to $r>L$, where it includes the future infinity ${\cal I}^+$ at $r \to \infty$.

In the new coordinates (\ref{boost}), the horizons satisfy 
\begin{equation}
  \frac{r'_H}{L'} = b_+ \frac{\cos \theta}{\cos \tau}- b_- \tan \tau  = \pm 1, \qquad \left(\f{r}{L}=\f{\cos\th}{\cos\t} \right).
\end{equation}
In terms of $r$, we have
\begin{equation}
 \frac{r}{L} = \frac{\pm 1 + b_- \tan \tau}{b_+}.
\end{equation}
At zero temperature  $\beta \to \infty$, $b_- \to 0, b_+ \to  1$ from \eqref{paramap}. Thus it coincides with the original horizons $r/L = \pm 1$, regardless of the value of $\tau$.

Consider the cosmic horizon choosing the positive sign. Since, for sufficiently sizable $\tau >0$, the location of the new horizon is,
\begin{equation} 
 \frac{r}{L} =  \frac{b_- \tan \tau +1}{b_+} >1, \quad |\theta| < \frac{\pi}{2},
 \end{equation}
which means that the new horizon is located at 
\begin{equation}
 r= L' =  \frac{b_- \tan \tau +1}{b_+}L
\end{equation}
which is  greater than $L$ for $\t$ near $\pi/2$.
That is, we have a growing cosmic horizon for nonzero $\beta$.

To see that the horizon of the black hole shrinks, we note that near $\t\rightarrow \pi/2$, the dilaton $\Phi$ diverges and changes sign at $\theta=\theta_0$, where
\begin{equation}
	\cos\theta_0 = \frac{b_-}{b_+} = \frac{ b^2 -1}{b^2+1}. 
\end{equation} 
So at $\t=\pi/2$, the range of $\th$ for singularity is ($\th_0,2\pi-\th_0)$, for which the dilaton $\Phi$ is negative. The future event horizon of the black hole is the null lines extended from the end points of the interval. 
The black hole horizon is measured at $\theta' = \pi+\theta$, thus
\begin{equation}
 \left|\f{r}{L}\right| = \left |\frac{-b_- \tan \tau+1}{b_+} \right|,
\end{equation}
which is less than 1 for sizable $\tau$. Thus we have a shrinking black hole horizon.

We have mentioned that as the CFT temperature increases, the causal structure changes and the black hole interior grows as in \autoref{backreactedds}. However, the black hole apparent horizon remain at $(\t,\th)=(0,\pi)$ and the dilaton value at that point decreases over temperature, corresponding to the evaporation of the black hole. 

On the other hand, the cosmological apparent and event horizon is moved to $(\t_0,0)$ near $\t=\pi/2$. It seems that the future light cone of the cosmological horizon gets smaller in the Penrose diagram. But the physical size of the horizon measured by the dilaton shows that it does not decrease. 

\subsection{Bousso--Hawking process}

The black hole that we have seen thus far is an eternal black hole in equilibrium at finite temperature $T=1/(\beta L)$. In the zero temperature limit $\beta \to \infty,$ and even if there is no temperature difference between the black hole and the de Sitter space $B \to \infty$, there is no radiation from the black hole and only the background radiation of dS$_2$ at $T_{\rm GH}$.

Ultimately, we are interested in the description of an evaporating black hole in de Sitter space and the emergence of pure dS$_2$ at the end of evaporation. We assume adiabatic evolution.
The BKU backreacted solution can be thought of as a local and instantaneous description of the black hole, which is in thermal equilibrium at a given CFT temperature. As time goes on, the black hole evaporates and the CFT temperature also increases so we may regard it as a parameter playing the role of time evolution. More realistic description of the decay of AdS black hole is done in \cite{Almheiri:2019psf}.

Another remark is that after the evaporation, the resulting geometry should be pure dS$_2$. Although the Penrose diagram changes due to backreaction, near the future infinity $\cal{I}^+$ we can rotate the coordinates by \eqref{boost} and transform the metric and dilaton into the Milne form \eqref{Milne}. This is consistent with the boundary condition \eqref{dilatonsl}.  So one can glue new dS$_2$ space along the future boundary of the existing one.  However, for an observer to be able to travel into the new de Sitter space within finite time, a subtlety arises. We shall come back to this issue in Section. \ref{sec:multiversesol}.

\section{Entropy computation}

In this section we calculate the entropy of a region in the future infinity $\cal{I}^+$ during the evaporation process. We consider limits of low, high and around the Gibbons--Hawking temperature. Islands appear in all cases. Around the critical temperature, we find the ``Page length'' $\th_R$, which corresponds to the size of the radiation-collecting region. In the high temperature approximation, the ``Page temperature'' is computed.

\subsection{The von Neumann entropy \label{sec:fintempentropy}}

To track the finite temperature effect of the black hole, we consider the ``Hartle--Hawking'' (HH) vacuum in a ``Schwarzschild''-like coordinate, in which the energy-momentum tensor is (\ref{flatt}). A stationary observer in the ``Schwarzschild''-like coordinate $x^{\pm}=L(\t\pm\th)$ observes the thermal radiation from the black hole. To calculate the entropy, we transform the coordinates such that the expectation value of the energy-momentum tensor (\ref{flatt}) vanishes
\begin{equation} \label{HHvac}
  \langle HH | \tau_{\pm \pm}(y^\pm) |HH \rangle = 0. 
\end{equation}
This is different from Bunch--Davies vacuum, whose temperature is related to the curvature of de Sitter space. 
Henceforth, the braket $\langle  \,\cdot \,\rangle$ denotes the expectation value for this Hartle--Hawking vacuum.

The following transformation 
\begin{equation} \label{HHtransf}
	y^{\pm} = \exp\left(-\frac{2 \pi}{BL} x^{\pm} \right), 
\end{equation}
with a notation of relative temperature $T_{r} = 1/(BL)$,
\begin{equation} \label{Trelative}
 T_r \equiv \frac{1}{BL} \equiv  \sqrt{\frac{1}{(\beta L)^2} - \frac{1}{(2\pi L)^2}} = \sqrt{T^2 - T^2_{\rm GH}},
\end{equation}
gives a vacuum state via
\begin{equation}
	\left(\f{\p y^{\pm}}{\p x^{\pm}}\right)^2 \langle  \t_{\pm\pm}(y^{\pm}) \rangle  = \langle  \t_{\pm\pm}(x^{\pm}) \rangle+\f{c}{24\pi}\{y^{\pm},x^{\pm}\},
\end{equation}
where  $\{y^{\pm},x^{\pm}\}= \f{y^{\pm'''}}{ y^{\pm'}} - \f{3}{2}\left(\f{y^{\pm''}}{ y^{\pm'}}\right)^2 = -\f{2\pi^2}{B L^2}$, with $y^{\pm'}$ denoting the derivative with respect to $x^\pm$. The metric \eqref{dsmet} is written in $y^{\pm}$ coordinates,
\begin{equation}
	ds^2= -\Omega(y)^{-2}dy^+dy^-, \qquad \Omega(y)^2 = \left(\f{2\pi}{BL}\right)^2y^+y^-\cos^2\left(\f{-B}{4\pi}\log y^+y^-\right).
\end{equation}
At zero temperature $T=0$, this reduces to the same transformation as in \cite{Hartman:2020khs}.

The von Neumann entropy is given by the single interval entropy of 2d CFT in a flat space together with a Weyl transformation by $\Omega(y)$ \cite{Calabrese:2004eu,Calabrese:2009qy}
\begin{align}\label{Smatfirst}
		S_{\text{mat}}(x_1,x_2) &= \f{c}{6}\log\left[\f{- (y_1^+-y_2^+)( y_1^--y_2^-)}{\e^2_{\text{uv}}\Omega(y_1)\Omega(y_2)}\right] \nn\\
		&= \f{c}{6}\log \left[\f{B^2L^2( \cosh \f{2\pi}{B}(\th_{1}-\th_2)-\cosh  \f{2\pi}{B}(\t_{1}-\t_2) )}{2\e_{\text{uv}}^2\pi^2\cos\t_1\cos\t_2} \right]\nn\\
		&= \f{c}{6}\log \left[\f{B^2L^2\left(\sinh\f{\pi}{BL}\D x^+\sinh\left(-\f{\pi}{BL}\D x^-\right)\right)}{\e_{\text{uv}}^2\pi^2\cos\left(\f{x_1^++x_1^-}{2L}\right)\cos\left(\f{x_2^++x_2^-}{2L}\right)}\right],
\end{align}
where we defined $\Delta x^\pm = x_1^\pm-x_2^\pm$.
We mostly use the global coordinates, so we view the formula as a function of $x^\pm$. One may note that the entropy formula in \cite{Balasubramanian:2020xqf} can be reproduced from \eqref{Smatfirst} by taking the difference of $S_{\text{mat}}$ at large $T$ and at $T=0$, i.e. $B\to - 2\pi i$.

When the relative temperature $T_r$ is positive, in the absence of the warp factor, the entropy formula is periodic under the imaginary time 
\begin{equation}
  \tau \simeq \tau + i B,
\end{equation}
which is the property of a thermal two-point correlation function \cite{Gibbons:1977mu}. Indeed, for high temperature $B \to \beta$ and this is the desired property of the black hole at temperature $T=1/(\beta L)$. For $T \le T_{\rm GH}$, $T_r$ becomes imaginary, making the entropy non-thermal. It implies that, if the black hole temperature $T$ is lower than the background temperature $T_{\rm GH}$ of de Sitter space, there is no thermal radiation from the black hole. We assume that the {\em absolute} temperature $T$ is the equilibrium temperature of the black hole and the CFT.

\subsection{Entropy with islands}

We use the matter entropy at finite temperature \eqref{Smatfirst} for an inflating region $R = \{x'_{R},x_R\}$ near the future infinity $\cal{I}^+$, depicted in \autoref{backreactedds} (we denote the left and right end of the interval by primed and unprimed coordinates, respectively). The region $R$ is in the asymptotically flat part of the geometry.
For an island located in the region $I=\{x'_{I},x_{I}\}$, the generalized entropy is given by
\begin{equation}
   S_{\text{gen}}(I\cup R)   = \frac{A(\partial I)}{4G_N} + S_{\text{mat}}(x_{I},x'_{R})+ S_{\text{mat}}(x_R,x'_{I})  ,
\end{equation}
up to UV counterterms, and the area of the island boundary $A(\partial I)$ is given by the value of the dilaton at those points $\phi(x_{I})$ and $\phi(x'_{I})$.
The arguments of $S_{\text{mat}}$ goes from the left to right end point of the intervals, as defined in \eqref{Smatfirst}.
This comes from the fact that the whole universe is pure and the corresponding entropy is zero, so that
$$
 S_{\text{gen}}(I\cup R)  = S_{\text{gen}}\left((I \cup R)^c \right), 
$$
in a Cauchy slice including the intervals $I$ and $R$.
Here an important assumption is that the cross-ratio of the relative positions of points are fixed so that the entropy is expressed in terms of the sum of the two-point correlation functions of the CFT \cite{Calabrese:2004eu}.

For simplicity, we take the region $R$ in a symmetric manner such that $\t_1=\t_2=\t_R$ and $\th_1=-\th_2=\th_R$. Also we require $\t_R$ close to $\pi/2$ so that $R$ becomes a non-gravitating region. Then, the matter entropy becomes
\begin{equation}\label{Smat}
	S_{\text{mat}}(R) = \f{c}{3}\log\left[\f{B L}{\pi \e }\sinh\f{2\pi}{B}\th_R \right] - \f{c}{3}\log(\cos\t_R ).
\end{equation}
And the generalized entropy is
\begin{align}\label{Sgen}
	S_{\text{gen}}(I\cup R) &= \f{\phi_0}{2G_N}+\f{\Phi(\t_I,\th_I)}{2G_N} + \f{c}{3}\log\left[\f{B L}{\pi \e }\sinh\f{\pi}{B}(\t_I-\t_R+\th_I-\th_R)\right] \\
	&~~~ + \f{c}{3}\log\left[\f{B L}{\pi \e }\sinh\f{\pi}{B}(-\t_I+\t_R+\th_I-\th_R)\right] - \f{c}{3}\log\big[\cos\t_I\cos\t_R\big], \nn
\end{align} 
where  the backreacted dilaton $\Phi(\t_I,\th_I)$ is given by \eqref{dilatonf}. 

The entanglement entropy is found by minimizing the generalized entropy with respect to $\theta_I$ and extremizing it with respect to $\tau_I$.
The condition for the derivative with respect to $\th_I$ to vanish is  
\begin{equation}\label{dwrtth}
	\p_{\th_I}S_{\text{gen}} = -\f{\bar{\phi}L}{2G_N}b_+\f{\sin\th_I}{\cos\t_I} +\f{\pi c}{3B}\left(\coth\f{\pi}{B}(\t_{IR}+\th_{IR})+\coth\f{\pi}{B}(-\t_{IR}+\th_{IR})\right) =0,
\end{equation}
where $\t_{IR} \equiv \t_I-\t_R$ and similarly for $\th_{IR}$.
The derivative of $S_{\text{gen}}$ with respect to $\t_I$ gives 
\begin{align}\label{dwrtt}
	\p_{\t_I}S_{\text{gen}} &= \f{\bar{\phi}L}{2G_N}\left(b_+\f{\cos\th_I\sin\t_I}{\cos^2\t_I} - \f{2}{\pi}b_-(\tan\t_I+\t_I\sec^2\t_I)\right) \nn\\
	& ~~~ +\f{\pi c}{3B}\left(\coth\f{\pi}{B}(\t_{IR}+\th_{IR}) - \coth\f{\pi}{B}(-\t_{IR}+\th_{IR})\right) + \f{c}{3}\tan\t_I=0.
\end{align}

\subsubsection{Low temperature limit}

For large $\b$, we have $B \to - 2\pi i$, and the temperature approaches zero. We summarize the results of Ref. \cite{Hartman:2020khs}. The matter entropy is
\begin{equation}
 S_{\text{mat,l}}(R) = \frac{c}{3} \log\frac{2L \sin \theta_R}{\epsilon} - \frac{c}{3} \log(\cos\tau_R).
\end{equation}
The generalized entropy is 
\begin{equation}
\begin{split} 
 S_{\text{gen,l}}(I\cup R) & = \frac{\phi_0}{2G_N} + \frac{\bar \phi L }{2 G_N} \frac{\cos \theta_I}{\cos\tau_I} \\
  &  + \f{c}{3}\log\left[\f{ 2L}{ \e }\sin\f{1}{2}(\t_I-\t_R+\th_I-\th_R)\right] \\
	& + \f{c}{3}\log\left[\f{2 L}{ \e }\sin\f{1}{2}(-\t_I+\t_R+\th_I-\th_R)\right] - \f{c}{3}\log\big[\cos\t_I\cos\t_R\big], 
\end{split}
\end{equation}
omitting the renormalization terms. In this case, the following coordinates in the hyperbolic patch simplify the analysis
\begin{equation}
 \tau = \tan^{-1} \left( \frac{\cosh X}{\sinh T} \right), \qquad \theta = \tan^{-1} \left(  \frac{\sinh X}{\cosh T}\right),
\end{equation}
by which we may define the parameters $(T_I,X_I)$ and $(T_R,X_R)$ in terms of $(\tau_I,\theta_I)$ and $(\tau_R,\theta_R)$, respectively. 

The entropy is minimized at $X_I = - X_R$. The width of the island decreases as the CFT interval gets smaller. Then  for small $T_R$, corresponding to the region $R$ near the future infinity or $\tau_R \to \pi/2$ where the gravity is weak, the entropy is extremized at $\sinh T_I = 3/(2\delta)$
with a small parameter,
\begin{equation} \label{delta}
 \delta \equiv \frac{ c G_N}{\bar \phi L} \ll 1,
\end{equation}
as explained at the end of Section \ref{sec:dilaton}.
The resulting von Neumann entropy is 
\begin{equation}
\begin{split}
 S_{\text{gen,l}}(\rho_R) &\simeq \frac{\phi_0 + \phi_I}{2G_N} + \frac{c}{6} \log \left( \frac{4 L^2(\phi_I-2cG_N/3)}{T_R^2 \epsilon^4(\phi_I+2cG_N/3)}\right),\\
 &\simeq \frac{\phi_0 - \bar \phi L}{2G_N} + \frac{c}{3} \log \left( \frac{2L}{T_R \epsilon^2}\right),
\end{split}
\end{equation}
with $\phi_I = -\bar \phi L\sqrt{1 +4\delta^2/9}.$ As the region $R$ gets wider beyond certain length, the island appears and the mixed states begin to be purified. In this sense, we may define the Page length $X_{R,\text{Page}}$ as that satisfying $S_{\text{gen,l}}(\rho_R)=S_{\text{mat,l}}(R)$, which is given by
\begin{equation} \label{LowTBeh}
\begin{split}
  \sinh X_{R,\text{Page}} 
 &\simeq   \frac{1}{\e} \sqrt\frac{\phi_I-2cG_N/3}{\phi_I+2cG_N/3}\exp\left(\frac{3(\phi_0+ \phi_I)}{2cG_N}\right)\\
&\simeq \frac{1}{\e}  \exp\left(\frac{3(\phi_0- \bar \phi L)}{2cG_N}\right),
\end{split}
\end{equation}
It corresponds to a very large region $R$.

\subsubsection{Critical temperature limit}

The temperature relation \eqref{Trelative} shows that the Gibbons--Hawking temperature plays the role of the critical one, above which an observer detects the finite temperature effect. The relative temperature measures the difference from the critical temperature, $B \to \infty$. 

In the critical temperature limit, the matter entropy \eqref{Smat} can be written as
\begin{equation}\label{lowtSmat}
	S_{\text{mat,c}}(R) = \f{c}{3}\log\f{2L\th_R}{\e} - \f{c}{3}\log(\cos\t_R),
\end{equation}
and the generalized entropy becomes
\begin{equation} \label{Sgencrit}
\begin{split}
 	S_{\text{gen,c}} (R \cup I)&= \f{\phi_0}{2G_N} + \f{\bar{\phi}L}{2G_N} \left[ b_+ \f{\cos\th_I}{\cos\t_I} -\frac{2}{\pi}b_- (\tau_I \tan \tau_I +1) \right ]\\
	&~ + \f{c}{3}\log\f{L^2(-\t_{IR}^2+\th_{IR}^2)}{\e^2\cos\t_I}-\f{c}{3}\log(\cos\t_R) \\
\end{split}
\end{equation}
The $\theta$-derivative \eqref{dwrtth} also simplifies to
\begin{equation} \label{Sgenctheta}
	\p_{\th_I}S_{\text{gen,c}}  \simeq -\f{\bar{\phi}L}{2G_N}b_+\f{\sin\th_I}{\cos\t_I} + \f{c}{3}\f{2\th_{IR}}{\th_{IR}^2-\t_{IR}^2}=0.
\end{equation} 
In this critical limit $B\to \infty$ or $\b\rightarrow 2\pi$, from \eqref{paramap}, we have
\begin{equation} \label{bApprox}
  b - \frac{1}{b} = \frac{\pi}{3} \delta,
\end{equation}
where the parameter $\delta$ is defined in \eqref{delta}.
This relation implies that the approximation $b \simeq 1 +\pi \delta/6$ holds.
Then with the definition below Eq. \eqref{boost}, $b_+=1$ and $b_- = \pi \delta/6$ to ${\cal O}(\delta^2)$ and the temperature dependence drops out in the condition \eqref{Sgenctheta}. 
In this case the backreaction should be very small, so we expand it in
 $\th_I$ and $\t_I$ around $\pi/2$, and also set 
$$\t_R=\f{\pi}{2}, \qquad \th_R \equiv \f{\pi}{2} -\e_\th,$$
 for the radiation collecting region $R$. The condition \eqref{Sgenctheta} becomes a quadratic equation in $\t_I$ and the solution gives the relations 
\begin{align}
	\t_I 
	&\cong \f{\pi}{2} + \left( \f{2}{3} \delta \pm\sqrt{1+\frac{4}{9}\delta^2}\right) (\theta_I-\theta_R),
\end{align}
for which we take the negative sign to have the interval inside the original dS, $\tau_I < \pi/2$. We have assumed $0<\theta_I-\theta_R \ll 1$ that we justify shortly.

In the $\t_I$-derivative of $S_{\text{gen}}$ \eqref{dwrtt}, 
we have 
\begin{equation}\label{lowtth}
\begin{split}
  \p_{\t_I}S_{\text{gen,c}}& \simeq  \f{\bar{\phi}L}{2G_N} \left( b_+ \f{\cos\th_I\tan\t_I}{\cos\t_I}
   - \frac{2}{\pi} b_- (\tan \tau_I + \tau_I \sec^2 \tau_I) \right) \\
  &~ - \f{c}{3}\f{2\t_{IR}}{\th^2_{IR}-\t_{IR}^2} + \f{c}{3}\tan\t_I=0.
\end{split}
\end{equation}
Here $\tan \tau_I \sim 1/\cos \tau_I \sim {\cal O}(\delta/\epsilon_\theta)$ are large.
Expanding it in $\th_I$ around $\pi/2$ to ${\cal O}(\delta^2)$ and using \eqref{lowttau} for $\t_I$, we find 
\begin{equation}
 \th_I  =\frac{\pi}{2}+\frac{2 \gamma^3 +9\cdot2^{1/3} \Gamma-6 \cdot 2^{2/3} \Gamma^{-1}\delta }{3 (3+4 \delta) } , 
\end{equation}
to order $\e_\th$ and $\delta$, with
\begin{equation}
\gamma \equiv (\pi  \delta-3 \epsilon_\theta)^{1/3}, \quad \Gamma \equiv 6^{-1/3} (3 \gamma^3+|3\gamma^3+26 \delta \epsilon_\theta|)^{1/3}.
\end{equation}
It further simplifies when $\gamma >0$, because we can neglect $ \delta \epsilon_\theta$ term in $\Gamma$ and approximate $\Gamma \simeq \gamma$.
Defining $\Delta \equiv \theta_I - \pi/2$, one can check that  $\Delta$ is roughly of order $\gamma$. So, $\Delta >0$ and the island region lies in $(\pi/2+\Delta,3\pi/2-\Delta) \subset (\pi/2,3\pi/2)$, see Fig. \ref{fig:eternalds2}. For $\gamma<0$ it may be that $\Delta<0$.
The smaller the region $\theta_R$ gets, the bigger the parameter $\epsilon_\theta$, hence the smaller the island. 

Using the result, we can also calculate
\begin{equation}\label{lowttau}
 \tau_I \cong \f{\pi}{2} + \left( \f{2}{3} \delta -\sqrt{1+\frac{4}{9}\delta^2}\right) (\Delta+\epsilon_\theta),
\end{equation}
as desired. The approximation is valid since $\theta_I-\theta_R = {\cal O}(\gamma) \ll 1$.

Plugging \eqref{lowttau} and \eqref{lowtth} into the generalized entropy \eqref{Sgencrit} at large $B$, we have 
\begin{align}\label{lowtSgen}
	S_{\text{gen,c}} (\rho_R)&\cong \f{\phi_0}{2G_N} -\f{\bar \phi L}{2 G_N} \left(1+\frac{\pi \delta}{6 \Delta}\right)+\f{c}{3}\log\left[\f{2L^2}{\e^2} \left( \f{\pi}{2}-\th_R\right) \right] -\f{c}{3}\log(\cos\t_R),
\end{align}
where we have restored $\theta_R$. Since $\delta/\Delta$ is of order $\delta^{2/3}$, one may neglect the term.
The Page length $\th_{R,\text{Page}}$ can be found by equating $S_{\text{gen,c}}(\rho_R)$ in \eqref{lowtSgen} and $S_{\text{mat,c}}(R)$ in \eqref{lowtSmat}. We find
\begin{equation}
	\th_{R,\text{Page}} \cong
	\f{\pi}{2}  \left[1-\f{\e\bar{\phi} }{cG_N}\exp \left(-\f{3(\phi_0-\bar \phi L)}{2cG_N}\right)\right].
\end{equation}
which is very close to but smaller than $\pi/2$, since $\phi_0 \gg \bar \phi L$ is assumed and the last term is exponentially suppressed. This behavior is similar to that in the low temperature case \eqref{LowTBeh}.

\subsubsection{High temperature limit}

\begin{figure}\begin{center}
		\begin{tikzpicture}[scale=0.85]
			\draw [gray, dashed] (0,0) -- (0,4);
			\draw[gray, dashed] (0,4) -- (4,4)
			;
			\draw[gray, dashed] (0,0) -- (4,0)
			;
			\draw [gray, dashed] (4,0) -- (8,0) -- (8,4) -- (4,4) -- (4,0);
			\draw (4.1,0) -- (4.1,4);
			\draw (-0.1,0) -- (-0.1,4);
			\draw (7.9,0) -- (7.9,4);
			\draw [decorate,decoration={zigzag,segment length=2.2mm, amplitude=0.6mm}] (-0.1,0) .. controls (2,0.3) .. (4.1,0);
			\draw [decorate,decoration={zigzag,segment length=2.2mm, amplitude=0.6mm}] (-0.1,4) .. controls (2,3.7) .. (4.1,4);
			\draw  (4.1,0) .. controls (6,0.3) .. (7.9,0);
			\draw  (4.1,4) .. controls (6,3.7) .. (7.9,4);
			\draw [very thick, blue] (4.8,3.8) .. controls (6,3.6) .. (7.2,3.8)
			node[midway, below] {$R$};
			\draw [very thick, magenta] (0.8,3.6) .. controls (2,3.5) .. (3.2,3.6)
			node[midway, below] {$I$};
			\draw (-0.1,0) -- (1.9,2) -- (-0.1,4);
			\draw (4.1,0) -- (2.1,2) -- (4.1,4);
			\draw (4.1,4) -- (6,2.1) -- (7.9,4);
			\draw (4.1,0) -- (6,1.9) -- (7.9,0);
			\filldraw[black] (2,2) circle (2pt);
			\filldraw[red] (6,2.1) circle (2pt);
			\filldraw[red] (6,1.9) circle (2pt);
			\draw (2,0.1)--(2,-0.1) node [below] {$\pi$};
			\draw (6,0.1)--(6,-0.1) node [below] {$0$};
		\end{tikzpicture}
		\hspace{0.7cm}
			\begin{tikzpicture}[scale=0.85]
			\draw [gray, dashed] (0,0) -- (0,4);
			\draw[gray, dashed] (0,4) -- (4,4)
			;
			\draw[gray, dashed] (0,0) -- (4,0)
			;
			\draw [gray, dashed] (4,0) -- (8,0) -- (8,4) -- (4,4) -- (4,0);
			\draw (4.5,0) -- (4.5,4);
			\draw (-0.5,0) -- (-0.5,4);
			\draw (7.5,0) -- (7.5,4);
			\draw [decorate,decoration={zigzag,segment length=2.2mm, amplitude=0.6mm}] (-0.5,0) .. controls (2,1) .. (4.5,0);
			\draw [decorate,decoration={zigzag,segment length=2.2mm, amplitude=0.6mm}] (-0.5,4) .. controls (2,3) .. (4.5,4);
			\draw  (4.5,0) .. controls (6,0.5) .. (7.5,0);
			\draw  (4.5,4) .. controls (6,3.5) .. (7.5,4);
			\draw [very thick, blue] (4.8,3.8) .. controls (6,3.4) .. (7.2,3.8)
			node[midway, below] {$R$};
			\draw [very thick, magenta] (1,3) .. controls (2,2.7) .. (3,3)
			node[midway, below] {$I$};
			\draw (-0.5,0) -- (1.5,2) -- (-0.5,4);
			\draw (4.5,0) -- (2.5,2) -- (4.5,4);
			\draw (4.5,4) -- (6,2.5) -- (7.5,4);
			\draw (4.5,0) -- (6,1.5) -- (7.5,0);
			\filldraw[black] (2,2) circle (2pt);
			\filldraw[red] (6,2.5) circle (2pt);
			\filldraw[red] (6,1.5) circle (2pt);
			\draw (2,0.1)--(2,-0.1) node [below] {$\pi$};
			\draw (6,0.1)--(6,-0.1) node [below] {$0$};
		\end{tikzpicture}
	\end{center} 
	\caption{The left diagram shows the region $R$ and island $I$ at low temperature. As the CFT temperature increases, the Penrose diagram is deformed further and the island moves downward to the apparent horizon of the black hole as drawn in the diagram on the right.}
	\label{Fig:lowhigh} 
	\end{figure}
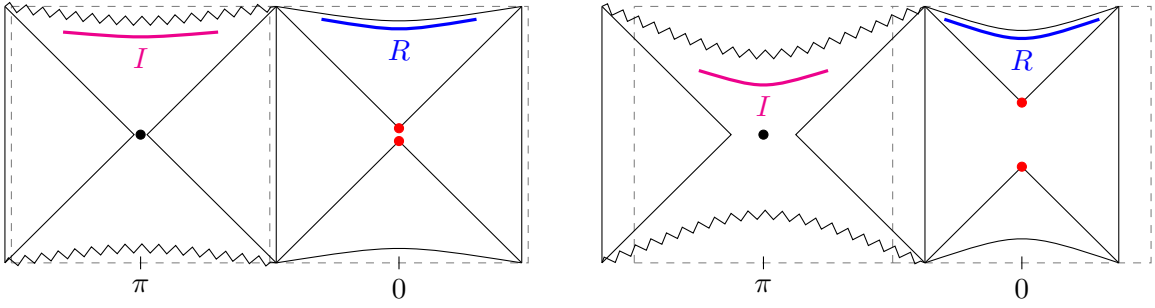

We now consider the case of high temperature, or small $\b$. In this limit, the absolute and relative temperatures are almost the same, $B \to \beta$. The matter and generalized entropy in \eqref{Smat} and \eqref{Sgen} can be approximated to
\begin{align}\label{Shigh}
	S_{\text{mat,h}} &= \f{2\pi c}{3 \beta}\th_R + \f{c}{3}\log\f{\b L}{2\pi \e} -\f{c}{3}\log(\cos\t_R), \nn\\
	S_{\text{gen,h}} &= \f{\phi_0}{2G_N} + \f{\bar{\phi}L}{4G_N}b\left(\f{\cos\th_I}{\cos\t_I} - \f{2}{\pi}(\t_I\tan\t_I+1)\right) +\f{c}{6} \nn\\
	& + \f{2\pi c}{3\b}\th_{IR} + \f{2c}{3}\log\f{\b L}{2\pi\e} - \f{c}{3}\log(\cos\t_I\cos\t_R).
\end{align}
From \eqref{paramap}, the parameters become $b_+ \cong b_- \cong \frac12 b\cong \f{2\pi^3cG_N}{3\bar{\phi}L}\f{1}{\b^2}$ for small $\b$. Since it appears only in the dilaton and is proportional to $b \propto 1/\b^2$, the dilaton dominstes the entropy $S_{\text{gen,h}}$. The $\theta_I$-extremization condition \eqref{dwrtt} can be expanded in $\b$ 
\begin{equation}
 \f{\bar{\phi}L b}{4G_N}\left( \cos \theta_I \sin \tau_I - \frac{2}{\pi}(\sin \tau_I \cos \tau_I + \tau_I) \right) + \frac{c}{3} \sin \tau_I \cos \tau_I = 0.
\end{equation}
and it is easy to see that $\t_I=0$ is a solution. The $\tau_I$-minimization condition \eqref{dwrtth} reduces to
\begin{equation}
	\f{\bar{\phi}L}{4G_N}b\left(-\f{\sin\th_I}{\cos\t_I}\right)+\f{2\pi c}{3\b}=0,
\end{equation}
and assuming $\th_I \approx \pi$, we find that 
\begin{equation}
	\th_I \cong \pi - \f{2\b}{\pi^2}.
\end{equation}
The size of the island is not sensitive to the region $R$ in the high temperature limit.
The location of $(\t_I,\th_I)$ shows that the island shrinks towards the apparent horizon of the blackhole, $(0,\pi)$, as temperature increases. See \autoref{Fig:lowhigh}. Using $\t_I$ and $\th_I$, we obtain
\begin{equation}
	S_{\text{gen,h}} = \f{\phi_0}{2G_N} - \f{\pi^2c}{3\b^2}(\pi+2) +\f{c}{6} +\f{2\pi c}{3\b }(\pi-\th_R) + \f{2c}{3}\log\f{\b L}{2\pi\e} - \f{c}{3}\log\cos\t_R.
\end{equation}

The Page temperature \cite{Page:1993wv} is computed by setting the above entropy equal to the matter entropy in \eqref{Shigh}. 
\begin{equation} \label{TPagehighT}
	T_{\text{Page}} \cong \sqrt{\f{3\phi_0}{2(\pi+2)\pi^2cG_N}} + \f{\pi-2\th_R}{(\pi+2)\pi}.
\end{equation}

\subsubsection{Intermediate temperature and the Page curve}

It is likely that the Page temperature is in the intermediate range. 
For this, we resort to numerical computation, to calculate the von Neumann entropy.

\begin{figure}
\begin{center}
\includegraphics[scale=1]{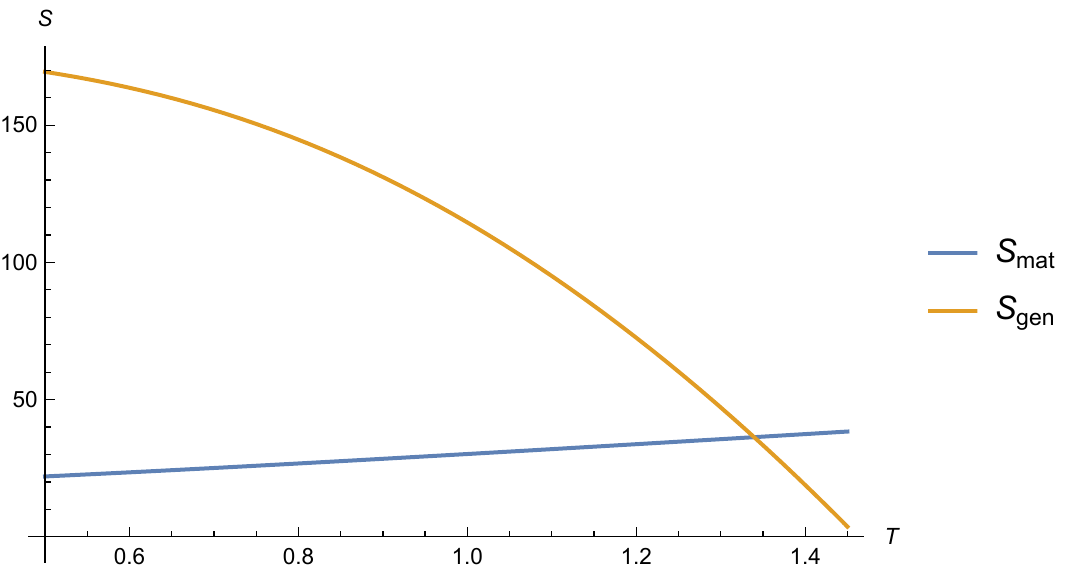}
\caption{Page curve for the evaporating black hole in dS$_2$ for a choice of parameters. The temperature of the black hole plays the role of time. \label{fig:pagecurve} }
\end{center}
\end{figure}

A typical page curve is drawn in \autoref{fig:pagecurve} with benchmark values $\phi_0=100,c=10,\bar \phi = 50,L=1,\theta_R=1,G_N=1/4,\epsilon=0.01$, in accord with the parameter ranges explained at the end of Section \ref{sec:backreactedsol}. 
The matter entropy \eqref{Smat} increases as temperature increases, reflecting that of radiation from the decaying black hole.
It is clear that for high enough temperature, the entanglement entropy \eqref{Sgen} is dominated by the dilaton term, as explained below Eq. \eqref{Shigh}, and decreases over temperature. If we decrease $\phi_0/G_N$, the Page temperature decreases, similar to \eqref{TPagehighT}. This dominant behavior is insensitive to other parameters than the temperature, thus the general curve should have a similar form.

Interestingly, numerical analysis shows that the minimum of generalized entropy is obtained for very small $\tau_I$ of order $0.1$ over large ranges of temperatures ${\cal O}(1)< \beta <{\cal O}(100)$. Assuming this, the extremization condition \eqref{dwrtt} is approximated as
\begin{equation}
 \tau_I \simeq  \frac{\f{\pi c}{3B}\left( \coth\f{\pi}{B}(-\frac{\pi}{2}+\th_{I}-\th_R)-\coth\f{\pi}{B}(\frac{\pi}{2}+\th_{I}-\th_R) \right)}{ \frac{\bar \phi L}{4 G_N}\left(b_+ \cos \theta_I-\frac{4}{\pi}b_- \right) - \frac{c}{3} }.
\end{equation}

\section{The de Sitter multiverse}

We consider multiple copies of global de Sitter spaces and calculate the entropy. Also, we address a question that to what extent the island formula is local. We expect it natural that a state in each individual de Sitter space should be pure even if multiple de Sitter spaces are generated from one de Sitter space; otherwise a wavefunction in one de Sitter space should be entangled with one in another space. Other possibilities are discussed in \cite{Levine:2022wos}. See also \cite{Choudhury:2020hil, Geng:2021wcq, Yadav:2022jib}.

\subsection{Multiverse spaces} \label{sec:multiversesol}

The BKU backreacted geometry, considered in Section \ref{sec:backreactedsol}, is characterized by the dilaton \eqref{dilatonf} that is periodic under $\theta \simeq \theta + 2\pi$. We may extend it to the multiple covers by allowing multiple periodicity $2\pi n$ with the same metric \eqref{dsmet}. This gives $n$ copies of de Sitter spaces having $2n$ black hole horizons and $2n$ cosmic horizons in the interval $\theta \in (0, 2 n \pi)$.

 Black holes and de Sitter spaces are symmetric with respect to the axes at $\theta = k \pi,$ with $k$ an integer. It corresponds to that each black hole backreacts at the same temperature. Otherwise, the black holes at different temperatures enforce the dilaton with different values of $b$ to be matched along some domain wall, which is not possible in the current setup of the BKU solution. We just leave a comment that different stress tensors for left and right black holes yields asymmetric radiation and corresponds to the asymmetric dilaton profile as in \eqref{dilaton}. 

\begin{figure} 
	\begin{center}
		\begin{tikzpicture}[scale=0.7]
			\draw (-4,4) -- (-4,2)--(12,2)--(12,4);
			\draw [decorate,decoration={zigzag,segment length=2.2mm, amplitude=0.6mm}] (-2,4) -- (2,4);
			\draw [decorate,decoration={zigzag,segment length=2.2mm, amplitude=0.6mm}] (6,4) -- (10,4);
			\draw (-4,4)--(-2,4);
			\draw (10,4)--(12,4);
			\draw (-4,2)--(-2,4)--(0,2)--(2,4)--(6,4)--(8,2)--(10,4)--(12,2);
			\draw (12,6)--(10,8);
			\draw [dashed] (2,2)--(2,4);
			\draw [dashed] (6,2)--(6,4);
			\draw [dashed] (-2,2)--(-2,4);
			\draw [dashed] (10,2)--(10,4);			
			\draw (2,2)--(2,0) -- (6,0) -- (6,2);
			\draw (2,0)--(6,0)
			node [midway, below] {${\cal I}_1^-$};			
			\draw (6,0) -- (2,4);
			\draw (6,4) -- (2,0);
			\draw [green] (2,4)--(6,4);
			\draw [green] (-4,4)--(-2,4);
			\draw [green] (10,4)--(12,4);
			\draw (-4,3) node[sloped, rotate=-50] {$\parallel$};
			\draw (12,3) node[sloped, rotate=-50] {$\parallel$};
			\draw [very thick, blue] (2.4,3.9)--(5.6,3.9) 
			node [midway, below] {$R_1$};
			\draw [very thick, blue] (10.4,3.9)--(12,3.9)
			node [midway, below] {$R_2$};
			\draw [very thick, blue] (-4,3.9)--(-2.4,3.9)
			node [midway, below] {$R_2$};
			\draw (2,4)--(2,8);
			\draw (6,8)--(6,4);
			\draw (2,8)--(6,8)
			node [midway, above] {${\cal I}_1^+$};
			\draw (2,4)--(6,8);
			\draw (2,8)--(6,4);
			\draw (-2,4)--(-2,8)--(-4,6)--(-2,4);
			\draw (-4,8)--(-2,8)
			node [midway, above] {${\cal I}_2^+$};
			\draw [dashed] (-4,4)--(-4,8);
			\draw (10,4)--(12,6)--(10,8)--(10,4);
			\draw (10,8)--(12,8)
			node [midway, above] {${\cal I}_2^+$};
			\draw [dashed] (12,4)--(12,8);
		\end{tikzpicture}
	\end{center} 
	\caption{Schematic diagram of two black holes generated at $\t=0$ from pure dS. A topology change of the spatial section from an interval to a circle occurs with two copies of global de Sitter spaces by the GPBH mechanism. After the black holes evaporate completely, two disjoint dS spaces emerge. We excise small regions at the spacelike interfaces of de Sitter spaces and glue them together along the green lines so that an observer can move to a new de Sitter space within finite time.} 
	\label{fig:multids}
\end{figure}
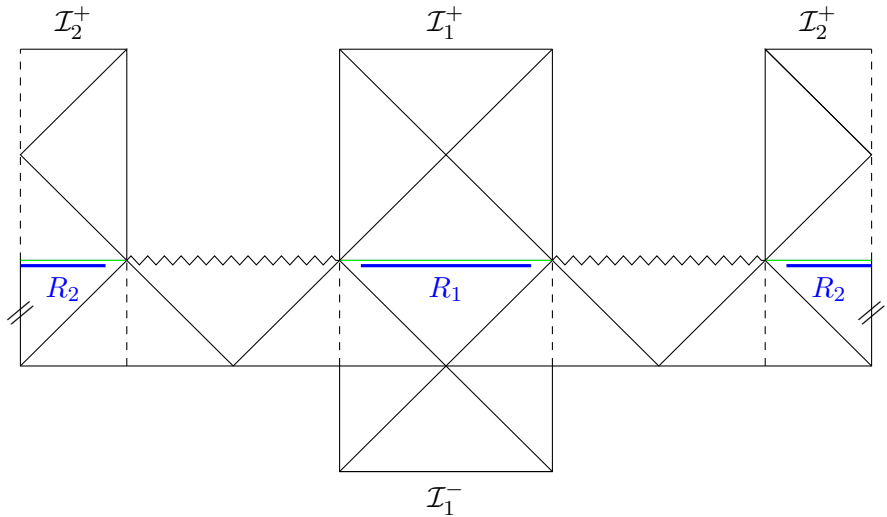

After black holes have evaporated, we are left with pure de Sitter spaces. To describe this, 
we attach newly created de Sitter spaces to the existing multiple de Sitter space. The Penrose diagram for a typical example is depicted in \autoref{fig:multids}.
For this, we may excise and glue the geometry as long as (i) the metric and (ii) the dilaton takes the same values at the interface \cite{Hartman:2020khs,Aguilar-Gutierrez:2021bns,Chen:2020tes}. Instead of attaching dS spaces, one can consider other spaces such as Minkowski or anti de Sitter space and glue them at the Milne ``expanding'' patch and/or even the similar ``crunching'' patch of the black hole region. 

We may attempt to glue two de Sitter spaces of different sizes by matching the metric in the global conformal coordinates,  
$$
 ds^2 = \begin{cases}
 \displaystyle \frac{L_e^2}{\cos^2\tau} (-d \tau^2 + d\theta^2), & ~~\tau < \tau_c, ~~~~(k-\frac12)\pi < \theta < (k+\frac12)\pi,\\
 \displaystyle   \frac{L_l^2}{\cos^2(\tau-\tau_c+\tau_l)} (-d \tau^2 + d\theta^2), & ~~\tau > \tau_c ,~~~~(k-\frac12)\pi < \theta < (k+\frac12)\pi,\\
  \end{cases}
$$
for a fixed integer $k$, satisfying 
$$ \frac{L_l^2}{ \cos^2 \tau_l }= \frac{L_e^2}{ \cos^2 \tau_c},$$
where $L_e$ and $L_l$ are radii of the dS spaces , $\t_l$ is a constant, and $\tau_c$ is the the position of a constant $\t$ cut of the two dS$_2$'s. The gravity action for each dS space should have different cosmological constants $2/L_e^2$ and $2/L_l^2$. However the warp factor of the neighboring region do not match at the interface $\theta= (k\pm 1/2)\pi$, for $\tau_c<\pi/2$.

Instead, we may cut the upper edge of the Milne patch of the existing de Sitter at $r'=1/\epsilon_{\rm s}$ with small enough $\epsilon_{\rm s}$, where $r'$ is the coordinate in the Milne patch after the boost transformation \eqref{boost} with the dilaton \eqref{dilatonsl}. This removes ${\cal I}^+$. Then we prepare another dS$_2$ whose opposite Milne patch is cut at the same distance, removing ${\cal I}^-$. By identifying the rest of coordinates, one can connect two de Sitter spaces. This is done in the exact same way as for the boundary of AdS to introduce a cutoff for UV divergence (The Penrose diagram for AdS is rotated by 90 degree in dS case). This also remedies the infinite time evolution problem between two consecutive de Sitter spaces, because the original boundary of the Milne patch ${\cal I}^+$ is not reachable at finite time from a point in the same de Sitter space.


\subsection{Multiple islands and locality}

We calculate von Neumann entropy for the $n$-copied multiverse.
Let us first consider the simplest case of single interval $n=1$. The generalized entropy \eqref{Sgen} depends on the difference between the nearest endpoints $\{ x^\pm_{I}, x'^\pm_{R}\}$ and  $\{x^\pm_R , x'^\pm_{I}\}$ as before, we the left and right end of the interval is denoted by primed and unprimed coordinates, respectively,
\begin{equation} \label{Sgenasymm}
 S_{\text{gen}}(I\cup R)  = \frac{\phi(x_{I})}{4G_N} + S_{\text{mat}}(x_{I},x'_{R})+ S_{\text{mat}}(x_R,x'_{I})+ \frac{\phi(x'_{I})}{4G_N}  .
\end{equation}
We assumed that $R$ is in the flat space so that there is no corresponding area (dilaton) terms. As long as the formula \eqref{Sgenasymm} is guaranteed to be correct, it is local in the sense that it does not depend on  the {\em details of the shape} of the intervals $I$ and $R$, but only on the endpoints of them. Formally the formula is sufficiently local and we heuristically use it to calculate the entropy in the multiple covers of the de Sitter space.

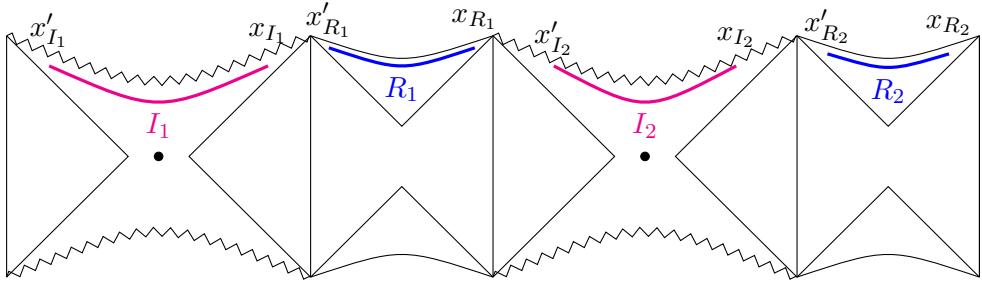
\begin{figure} \begin{center}
		\begin{tikzpicture}[scale=0.8]
			\draw (4.5,0) -- (4.5,4);
			\draw (-0.5,0) -- (-0.5,4);
			\draw (7.5,0) -- (7.5,4);
			\draw [decorate,decoration={zigzag,segment length=2.2mm, amplitude=0.6mm}] (-0.5,0) .. controls (2,1) .. (4.5,0);
			\draw [decorate,decoration={zigzag,segment length=2.2mm, amplitude=0.6mm}] (-0.5,4) .. controls (2,3) .. (4.5,4);
			\draw  (4.5,0) .. controls (6,0.5) .. (7.5,0);
			\draw  (4.5,4) .. controls (6,3.5) .. (7.5,4);
			\draw [very thick, blue] (4.8,3.8) node[above, black]{$x'_{R_1}$} .. controls (6,3.4) .. (7.2,3.8) node[above=1mm, black]{$x_{R_1}$}
			node[midway, below] {$R_1$};
			\draw [very thick, magenta] (0.2,3.5) node[above=1mm, black]{$x'_{I_1}$}  .. controls (2,2.7) .. (3.8,3.5) node[above=1.5mm, black]{$x_{I_1}$} 
			node[midway, below] {$I_1$};
			\draw (-0.5,0) -- (1.5,2) -- (-0.5,4);
			\draw (4.5,0) -- (2.5,2) -- (4.5,4);
			\draw (4.5,4) -- (6,2.5) -- (7.5,4);
			\draw (4.5,0) -- (6,1.5) -- (7.5,0);
			\filldraw[black] (2,2) circle (2pt);
			
			\draw (12.5,0) -- (12.5,4);
			\draw (15.5,0) -- (15.5,4);
			\draw (15.5,0) -- (15.5,4);
			\draw [decorate,decoration={zigzag,segment length=2.2mm, amplitude=0.6mm}] (7.5,0) .. controls (10,1) .. (12.5,0);
			\draw [decorate,decoration={zigzag,segment length=2.2mm, amplitude=0.6mm}] (7.5,4) .. controls (10,3) .. (12.5,4);
			\draw  (12.5,0) .. controls (14,0.5) .. (15.5,0);
			\draw  (12.5,4) .. controls (14,3.5) .. (15.5,4);
			\draw [very thick, blue] (13,3.7) node[above, black]{$x'_{R_2}$}  .. controls (14,3.4) .. (15,3.7) node[above=1mm, black]{$x_{R_2}$} 
			node[midway, below] {$R_2$};
			\draw [very thick, magenta] (8.5,3.5)  node[above, black]{$x'_{I_2}$}  .. controls (10,2.7) .. (11.5,3.5) node[above=1mm, black]{$x_{I_2}$} 
			node[midway, below] {$I_2$};
			\draw (7.5,0) -- (9.5,2) -- (7.5,4);
			\draw (12.5,0) -- (10.5,2) -- (12.5,4);
			\draw (12.5,4) -- (14,2.5) -- (15.5,4);
			\draw (12.5,0) -- (14,1.5) -- (15.5,0);
			\filldraw[black] (10,2) circle (2pt);
		\end{tikzpicture}
	\end{center} 
	\caption{Islands for the $n=2$ covering space. We may choose different CFT intervals $R_1,R_2$ that they may collect radiation from the adjacent black holes at a different rate. Corresponding islands $I_1$ and $I_2$ are formed. However, the entropy formula is formally local so that the extremization conditions determine the end points of the islands, neighboring an $R_i$.} \label{fig:twocovers}
\end{figure}

For later convenience, we define an entropy of symmetric interval such that 
\begin{equation} \label{Ssymm}
 \tilde S_{\text{gen}}(x_I,x_R) \equiv  \frac{\phi(x_{I})}{2G_N} + 2S_{\text{mat}}(x_{I},x_{R}).
\end{equation}
That is, the entropy in the case $x_R'=-x_R, x_I'=-x_I$. In the argument, we emphasized the coordinates of the endpoints of the intervals $I$ and $R$, instead of the set $I \cup R$. 

The extrimization conditions of $ S_{\text{gen}}(I\cup R)$ for $\theta_I, \tau_I, \theta_I', \tau_I'$ can be written in terms of the symmetric entropy, $\tilde{S}_{\text{gen}}$, as 
\begin{align}
 \frac {\partial  S_{\text{gen}}(I\cup R)}{\partial \tau_I} &=  \frac{1}{4G_N}\frac{\phi(x_{I})}{\partial \tau_I} +\frac{\partial S_{\text{mat}}(x_{I},x'_{R})}{\partial \tau_I} = \frac{1}{2} \frac {\partial   \tilde S_{\text{gen}}(x_I,x_R')}  {\partial \tau_I}=0, \\
 \frac {\partial  S_{\text{gen}}(I\cup R)}{\partial \theta_I} &=  \frac{1}{4G_N}\frac{\phi(x_{I})}{\partial \theta_I} +\frac{\partial S_{\text{mat}}(x_{I},x'_{R})}{\partial \theta_I} = \frac{1}{2} \frac  {\partial   \tilde S_{\text{gen}}(x_I,x_R')}   {\partial \theta_I}=0, \\
  \frac {\partial  S_{\text{gen}}(I\cup R)}{\partial \tau_I'} &=  \frac{1}{4G_N}\frac{\phi(x'_{I})}{\partial \tau_I'} +\frac{\partial S_{\text{mat}}(x'_{I},x_{R})}{\partial \tau'_I} = \frac{1}{2} \frac  {\partial   \tilde S_{\text{gen}}(x'_I,x_R)}   {\partial \tau_I'}=0, \\
 \frac {\partial  S_{\text{gen}}(I\cup R)}{\partial \theta'_I} &=  \frac{1}{4G_N}\frac{\phi(x'_{I})}{\partial \theta_I'} +\frac{\partial S_{\text{mat}}(x'_{I},x_{R})}{\partial \theta'_I} = \frac{1}{2} \frac  {\partial   \tilde S_{\text{gen}}(x'_I,x_R)}  {\partial \theta_I'}=0.
 \end{align}
The first two equations do not affect the last two and vice versa. In other words, the extrimization of the island boundary separately takes place for each cross-interval.
 
In the end, the extremized entropy becomes the average of the symmetric ones
\begin{equation}
 \text{ext}_I S_{\text{gen}}(I\cup R) = \text{ext}_I \frac{1}{2} \left( \tilde S_{\text{gen}}(x_I,x_R') 
 +  \tilde S_{\text{gen}}(x_I',x_R) \right).
\end{equation}
Therefore, for asymmetric $R$ where $x'_{R} \ne - x_{R}$, we may obtain the entropy by averaging the symmetric ones. In the symmetric case of $x'_{R} = - x_R$, we naturally have symmetric island $x'_{I}=-x_I$ so that two of the above are nontrivial.

Now consider the case of two covers $n=2$. The situation is depicted in \autoref{fig:twocovers}. 
We take two intervals $R_1 = (x'_{R_1},x_{R_1})$ and $R_2= (x'_{R_2},x_{R_2})$ in the Milne patches. In this case each observer collects the radiation from both adjacent, the left- and the right-, black holes. The islands $I_1$ and $I_2$ can form inside the black holes, as in \autoref{fig:twocovers}.
The generalized entropy is
\begin{equation}\begin{split}
 S_{\text{gen}} (I_1 \cup R_1 \cup I_2 \cup R_2)& =  \frac{\phi(x_{I_1})}{4G_N} +S_{\text{mat}}(x_{I_1},x'_{R_1}) + S_{\text{mat}}(x_{R_1},x'_{I_2})+ \frac{\phi(x'_{I_2})}{4G_N}  \\
&  + \frac{\phi(x_{I_2})}{4G_N}  +S_{\text{mat}}(x_{I_2},x'_{R_2}) + S_{\text{mat}}(x_{R_2},x'_{I_3})+ \frac{\phi(x'_{I_3})}{4G_N},
 \end{split}
\end{equation}
where $x'_{I_3}$ is identified with $x'_{I_1}$.

The radiation collected in the region $R_1$ comes from both the black holes containing $I_1$ and $I_2$, respectively. However, the entropy is only dependent on the cross-interval $x_{I_1} - x'_{R_1}, x_{R_1}- x'_{I_2}$. 
Consequently, the radiation falling in each $R_1$ and $R_2$ is purified separately. Each symmetric entropy (\ref{Ssymm}) provides a building block of the interval, containing half $R_i$ and its neighboring half $I_i$.

This possibility of locality in purification can be understood in the following way.
At a given interval $R_i$, only the nearest black holes affect the entropy. If black holes are sufficiently separated in the spacelike direction, they should be causally disconnected. Of course, we assume that the intervals $R_i$ and $I_{i}, I_{i+1}$ are sufficiently large so that the entropy is dominated by two-point correlation functions, not the higher. Another point to mention is that in the Penrose diagram, the bifurcate horizons disappear due to the growth of black hole inside. This means that left and right one sided black holes are causally disconnected. 

We may consider $R_2 \to \emptyset$ limit. Shrinking $R_2$ makes $I_1$ and $I_2$ smaller at the same time, moving $x_{I_2}$ to the left and $x'_{I_1}$ to the right, while leaving other coordinates $x'_{I_2}$ and $x_{I_1}$ intact. At high temperature, these points approach to the apparent horizons of the corresponding black holes. One must be careful, however, that the intervals $(I \cup R)^c$ should be kept small enough that the two-point correlation function is dominant in the entropy formula. With the vanishing $R_2$, we still have the two islands, because the local extremization condition for the {\em ``the radiation collector'' $R_1$} does not necessarily imply that the islands $I_1$ and $I_2$ are connected, but only provides information on two endpoints near $R_1$. As emphasized in \cite{Gibbons:1977mu}, the notion of thermal effect or entanglement is {\em observer-dependent}, so that it is meaningless to assume an ``observation'' without specifying CFT that is in thermal contact.

The generalization for arbitrary number of universes is straightforward. The entropy for $n$-copy multiverse is
\begin{equation}
\begin{split} \label{Sgenforn}
 S_{\text{gen}} \left(\bigcup_{k=1}^n(I_k \cup R_k) \right)&=  \sum_{k=1}^{n-1} \left[ \frac{\phi(x_{I_k})}{4G_N} +S_{\text{mat}}(x_{I_k},x'_{R_k}) + S_{\text{mat}}(x_{R_k},x'_{I_{(k+1)}})+ \frac{\phi(x'_{I(k+1)})}{4G_N}  \right] \\
 & =  \sum_{k=1}^{n-1} \f{1}{2}\left[\tilde S_{\text{gen}}(x_{I_k},x'_{R_k}) + \tilde S_{\text{gen}}(x_{R_k},x'_{I_{(k+1)}}) \right],
 \end{split}
\end{equation}
with the identification $x'_{I_n} \equiv x'_{I_1}$. Again the $\tilde S_{\text{gen}}$ is the symmetric entropy (\ref{Ssymm}). Once the dilaton values are calculated as in the next section, the extremization can be done in each $R_k$ as above.

\subsection{Finite temperature entropy} \label{sec:casimir}

The Casimir energy changes in the energy momentum tensor \cite{Aguilar-Gutierrez:2021bns,Levine:2022wos} due to the change in the $\th$ periodicity. The stress tensor of a thermal state with $2\pi n$ periodicity of $\th$ coordinate is given by 
\begin{equation}\label{nflatt}
	\t_{\pm\pm} = \f{c}{48\pi L^2}\left(\f{2\pi}{\b}\right)^2 - \f{c}{48\pi n^2L^2} .
\end{equation}   
That is, the Casimir energy is suppressed by the factor $n^2$ compared with \eqref{flatt}.
The energy momentum tensor in the dS$_2$ background is 
\begin{equation}
	\langle T_{\pm\pm} \rangle  = \f{c}{48\pi L^2}\left(\f{2\pi}{\b}\right)^2+ \f{c}{48\pi L^2}\left(1-\f{1}{n^2}\right).
\end{equation}
The dilaton takes the form
\begin{equation}\label{dilatonn}
	\Phi(\t,\th) = \a\f{\cos\th}{\cos\t} - K_n\left(\t\tan\t +1\right) +\f{cG_N}{3},
\end{equation}
with 
\begin{equation}
	K_n=\f{cG_N}{3}\left(\f{2\pi}{\b}\right)^2 + \f{cG_N}{3}\left(1-\f{1}{n^2}\right).
\end{equation} When $n=1$, $K_n$ reduces to $K$ defined in Eq. \eqref{paramap}. Imposing the boundary condition at $\t=\pi/2$ that the dilaton describes asymptotically pure de Sitter space implies
\begin{equation}\label{nparamap}
	K_n = \f{\bar{\phi}L}{\pi}\left(b-\f{1}{b}\right) ,\qquad \a = \f{\bar{\phi}L}{2}\left(b+\f{1}{b}\right),
\end{equation}
where $b$ is the parameter introduced in \eqref{dilatonsl}.

To compute the entropy, we perform a similar transformation as \eqref{HHtransf}
\begin{equation}
	y^{\pm} = \exp\left(-\frac{2 \pi}{B_n L} x^{\pm} \right), \qquad \frac{2\pi}{B_n} = \sqrt{\left(\f{2\pi}{\b}\right)^2-\f{1}{n^2}},
\end{equation} 
so that the energy momentum tensor vanishes in $y_n^{\pm}$ coordinates. Accordingly, the Hartle--Hawking vacuum is defined. The matter entropy for an interval $R_i=(x^{\pm}_{1,R_i},x^{\pm}_{2,R_i})$ is same as before but with $B_n$, 
\begin{align}
	S_{\text{mat}, n}(R_i) &= \f{c}{6}\log \left[\f{2L^2}{\e_{\text{uv}}^2}\left(\sinh\f{\pi }{B_n L}\D x^+_{R_i}\sinh\left(-\f{\pi}{B_n L}\D x^-_{R_i}\right)\right)\right] \nn\\
	& - \f{c}{6}\log\left[\left(\f{2\pi}{B_n}\right)^2\cos\left(\f{x_{1,R_i}^++x_{1,R_i}^-}{2L}\right)\cos\left(\f{x_{2,R_i}^++x_{2,R_i}^-}{2L}\right)\right]
\end{align}
Both small $\b$ limit and large $n$ limit lead to the same entropy formula as in the previous section, exception that the relative temperature now satisfies $T_r^2 = T^2 - (T_{\rm GH}/n)^2$. Also, in the large $\b$ limit it reduces to well-known multiverse entropy formula in \cite{Aguilar-Gutierrez:2021bns}. We are interested in the entropy of  $n$ regions in inflating patches as in \autoref{fig:multids}. Thus $S_{\text{total}} = \sum_{i=1}^nS_{\text{mat}, n}(R_i)$.

The generalized entropy in $n$ black hole case is given in \eqref{Sgenforn}. We have taken all $R_i$'s as $n$ copies of a single region. 
Since $B_n \rightarrow \beta$ at high temperature, we just have $n$ identical islands of previous section.
\begin{equation}
	\t_{I_i} \cong 0, \qquad \th_{I_i} = (2i-1)\pi - \f{2\b}{\pi^2} \qquad (i=1,...,n)
\end{equation}  
The Page temperature is 
\begin{equation}
	T_{\text{Page}} \cong \sqrt{\f{3\phi_0}{2(\pi+2)\pi^2cG_N}} + \f{\pi-2\th_{R_i}}{(\pi+2)\pi},
\end{equation}
for each black hole.

\section{Discussion}

We have considered the Ginsparg--Perry--Bousso--Hawking mechanism, which provides a good scenario for proliferating de Sitter spaces. Gravitational instanton effect can attach a handle to the spatial section of de Sitter space, connecting the two poles of dS to form the Nariai limit of the Schwarzschild de Sitter space. In two-dimensions, by coupling the JT gravity to a conformal matter, we let the black hole evaporate at finite CFT temperature above the Gibbons-Hawking temperature, using the Balasubramanian--Kar--Ugajin solution.

The BKU solution includes the backreaction from the semiclassical expectation value of the stress tensor for a thermal state in the global conformal coordinates of de Sitter space. We have computed the entanglement entropy of an interval in the expanding patch of dS$_2$ by a coordinate transformation to a vacuum state and demonstrated that it becomes zero at the end due to the purification of the state by the island. 

Then, the generation of multiple de Sitter spaces was considered and the entropy of them was investigated, including the change in the Casimir energy. We introduced multiple islands, where disconnected islands appear in the black hole inside  and showed that after the black holes have evaporated completely, the entropy of each new universe vanishes. It implies that each universe is locally pure and does not know the existence of other multiverses, though other possibilities exist. 

We close the discussion by comparing two versions of de Sitter multiverse evolution scenarios. 
In the Bousso--Hawking (BH) model \cite{Bousso:1998bn,Bousso:2002fq} the multiverse is described by fluctuation of Foruier modes along the $S^1$ direction, which is the spatial section of dS$_2$ space. That is, the dilaton is given as
\begin{equation}
 \phi = \frac{1}{\Lambda_2} \left( 1 - 2 \epsilon_{\rm BH} \sum_n \sigma_n(\tau) \cos n \theta    \right),
\end{equation}
with a small perturbation parameter $\epsilon_{\rm BH}$ and modified cosmological constant $1/\Lambda_2 = (1-N/(3 \Lambda^2))/\Lambda^2$. Here $N$ is the number of scalar fields which has thermal contact to the black hole. They contribute to the one-loop effects in the effective action and their energy momentum tensor gives rise to the fluctuation $\sigma_n(\tau)$. The profile $\sigma_n(\tau)$ is determined by the solution to the two-dimensional Einstein equation. The range of the parameter is limited as $\theta \in (0,2\pi)$, so the Casimir energy is always in the form \eqref{flatt}, regardless of the number of de Sitter spaces $n$. 

The Bousso--Hawking solution is very similar to the BKU solution in the zero temperature limit and for $n=1$. Indeed the classical solution matches $\sigma_1(\tau) = 1/\cos \tau$ and both solutions consider the fluctuation around it. However the classical solution of dilaton \eqref{dilaton1} can only describe the sourceless part. In the BKU solution, the constant energy momentum tensor from matter source modifies the dilaton but the modification is independent on $\theta$ and only depends on $\t$. Thus the higher Fourier mode in $\th$ is not possible in the BKU solution. 

For future directions, we mention that though we have not studied in this article, one may consider the entanglement between the disjoint de Sitter spaces. The entropy formula has been proposed in \cite{Balasubramanian:2020coy, Balasubramanian:2021wgd}. See also \cite{Geng:2020fxl,Geng:2021iyq}.
It would also be interesting to understand the application of the entanglement islands in the recent proposal of the de Sitter holography \cite{Susskind:2021dfc, Susskind:2021esx, Susskind:2021omt, Shaghoulian:2021cef, Shaghoulian:2022fop}.

\acknowledgments

This work is supported by the grant NRF-2018R1A2B2007163 of National Research Foundation of Korea.

\newpage
\providecommand{\href}[2]{#2}\begingroup\raggedright

\endgroup

\providecommand{\href}[2]{#2}\begingroup\raggedright
\bibliography{references}
\bibliographystyle{JHEP}
\endgroup

\end{document}